\documentclass[12pt,twoside,draft]{article}
\usepackage{amsmath}
\usepackage{amsfonts,amssymb,latexsym}
\usepackage{amscd}
\usepackage{eucal,mathrsfs}%,euscript}
\usepackage{amsthm}
\newtheorem{theorem}{\bf{Theorem}}[section] 
\newtheorem{corollary}[theorem]{\bf{Corollary}}

\newtheorem{proposition}[theorem]{\bf{Proposition}} 
 
\newtheorem{definition}[theorem]{\bf{Definition}}%[section]

\theoremstyle{remark}

\newtheorem{remark}{Remark}

\usepackage{graphics}
\numberwithin{equation}{section}
\def\Bbb{\mathbb}
\newcommand{\wick}[1]{\mathopen:\,\,#1\,\,\mathclose:} %% Wick-ordering
\newcounter{contador}

\def\logo{\raisebox{-10.5\p@}{\hb@xt@85\p@{\includegraphics{gft.eps}\hfil}}}
\def\un{1\kern-3pt \rm I}
\def\ptoday{{\ifcase\month 
\or January, \or February, \or March, \or April,\or May, 
\or June, \or July, \or August, \or September, \or October, 
\or November, \or December,\fi\ \number \year}}
\newcommand{\oR}{{\mathbb R}}

\newcommand{\oC}{{\mathbb C}}

\input epsf
\usepackage[dvips]{graphicx}
\def\dj{\hbox{d\kern-0.347em \vrule width 0.3em height 1.252ex depth
-1.21ex \kern 0.051em}}

\begin{document}
%%%%%%%%%%%%%%%%%%%%%%%%%%%%%%%%%%%%%%%%%%%%%%%%%%%%%%%%%%%%%%%%%%%%%%%%%%%%%%%%%%

\title{\rm On the Borchers Class of a Non-Commutative Field}

\author{\normalsize{\bf Daniel H.T. Franco}\footnote{On leave from the
Centro de Estudos de F\'\i sica Te\'orica, Belo Horizonte, MG,
Brasil.} \footnote{e-mail: dhtf@terra.com.br}\\
\\
{\normalsize {\em Universidade Federal do Esp\'\i rito Santo (UFES)}}\\
{\normalsize {\em Departamento de F\'\i sica -- Campus Universit\'ario de Goiabeiras}}\\
{\normalsize {\em CEP:29060-900 -- Vit\'oria -- ES -- Brasil}}}

\date{\ptoday}

\maketitle

\vspace{-1cm}

%\begin{center}
%\underline{\bf DRAFT VERSION}
%\end{center}

\begin{abstract}
In this paper, we arrive at the notion of equivalence classes of
a non-commutative field exploring some ideas by Soloviev to nonlocal
quantum fields. Specifically, an equivalence relation between non-commutative
fields is formulated by replacing the weak relative locality condition by a weak
relative asymptotic commutativity property, generalizing the notion of relative
locality proved by Borchers in the framework of local QFT. We restrict ourselves
to the simplest case of a scalar field theory with space-space non-commutativity.  
\end{abstract}


%newpage

%%%%%%%%%%%%%%%%%%%%%%%%%%%%%%%%%%%%%%%%%%%%%%%%%%%%%%%%%%%%%%%%
\section{Introduction}
%\hspace*{\parindent}
%%%%%%%%%%%%%%%%%%%%%%%%%%%%%%%%%%%%%%%%%%%%%%%%%%%%%%%%%%%%%%%%
At the present, a considerable effort has been made to clarify the structural
aspects of non-commutative field theories (NCFT) from an axiomatic standpoint.
The first paper within this context is due to \'Alvarez-Gaum\'e and
V\'asquez-Mozo~\cite{AGVM}. By modifying the standard Wightman axioms, using as guiding
principles the breaking of Lorentz symmetry down to the subgroup $O(1,1) \times SO(2)$,
which leaves invariant the commutation relations for the coordinate operators
$[\widehat{x}_\mu,\widehat{x}_\nu]=i \theta_{\mu\nu}$, and the relaxation of local
commutativity to make it compatible with the causal structure of the theory, described
by the light-wedge associated with the $O(1,1)$ factor of the kinematical symmetry group,
they have demonstrated the validity of the CPT theorem for NCFT with space-space
non-commutativity. More recently, Chaichian {\it et al.}~\cite{Cha1} have proposed new
Wightman functions as vacuum expectation values of products of field operators in the
non-commutative space-time. These Wightman functions involve the $\star$-product. In the
case of only space-space non-commutativity ($\theta_{0i}=0$), they have proved the CPT
and Spin-Statistics theorem, for the simplest case of a scalar field, using the
non-commutative form of the Wightman functions.  

As it was emphasized in~\cite{AGVM}, a source of difficulties in formulating NCFT which
satisfy the adapted axioms has been the very harmful UV/IR mixing, which is probably
the most surprising feature of these theories. The existence of hard infrared
singularities in the non-planar sector of the theory, induced by uncancelled quadratic
ultraviolet divergences, can result in two kinds of problems: they can destroy the
{\em tempered} nature of the Wightman functions and/or they can introduce tachyonic
states in the spectrum, so the modified postulate of local commutativity is not
preserved. In order to find a way of avoiding these problems, and due to the highly
nonlocal character of the commutation relations $[\widehat{x}_\mu,\widehat{x}_\nu]=
i \theta_{\mu\nu}$, one might want to consider other spaces of distributions. Having
this in mind, in~\cite{DCJhep}, we adopted the space of distributions which has been
explored some time by L\"ucke~\cite{Luc1}-\cite{Luc3} and Soloviev~\cite{Solo1}-\cite{Solo4}.
These authors have shown that one adequate solution to treat field theories with nonlocal
interactions, it is to take the fields to be averaged with test functions belonging to the
space ${\cal S}^0({\Bbb R}^n)$, consisting of the restrictions to ${\Bbb R}^n$ of entire
analytic functions on ${\Bbb C}^n$, whose Fourier transform is just the Schwartz space
${\mathscr D}({\Bbb R}^n)$ of $C^\infty$ functions of compact support. The space
${\cal S}^0({\Bbb R}^n)$ is the smallest space among the Gelfand-Shilov spaces~\cite{GeSh},
${\cal S}^\beta({\Bbb R}^n)$, where $0\leq \beta < 1$, which naturally allows us to treat
a theory of {\em nonlocalizable fields}. Elements in the dual space ${\cal S}^{\prime 0}$
of the space of entire functions are called analytic functionals. Because the elements in
${\cal S}^0$ are entire functions, the locality axiom cannot be formulated in the usual
way, {\em i.e.}, there is no sensible notion of support for distributions in
${\cal S}^{\prime 0}$. Nevertheless, Soloviev has shown that the functionals of this
class retain a kind of {\em angular localizability}, which ensures the
existence of minimal carrier cones of the distributions in ${\cal S}^{\prime 0}$, which
allows us handling the analytic functionals of class ${\cal S}^{\prime 0}$, in most
cases, as easily as tempered distributions. This replaces the notion of support for
nonlocalizable distributions and leads to a natural generalization of the local
commutativity. With this, we have been able of reaching a conclusion about the validity
of the CPT and Spin-Statistics theorems in NCFT, in the case of only space-space
non-commutativity~\cite{DCJhep}.

In this paper, following our previous paper~\cite{DCJhep}, we define what is
meant by an equivalence class of quantum fields for a non-commutative field
directly applying some ideas by Soloviev to nonlocal quantum fields in~\cite{Solo4}.
In ordinary field theories this was first described by Borchers~\cite{Bor,BEC}. We
restrict ourselves to the simplest case of a scalar field theory with space-space
non-commutativity.

The article is organized as follows. In Section 2, the necessary modifications of the
Wightman axioms to include the case of NCFT are described. In Section 3, we outline the
arguments that lead to a natural extension of the Borchers classes of quantum fields.
Section 4 contains our final considerations. An informal synopsis of some results by
Soloviev about carrier cones of analytic functionals is given in an Appendix.

%%%%%%%%%%%%%%%%%%%%%%%%%%%%%%%%%%%%%%%%%%%%%%%%%%%%%%%%%%%%%%%%%%%%%
\section{Modifying of the Wightman Axioms for a NCFT}
%\hspace*{\parindent}
%%%%%%%%%%%%%%%%%%%%%%%%%%%%%%%%%%%%%%%%%%%%%%%%%%%%%%%%%%%%%%%%%%%%%
We shall assume a NCFT with the field operators $\varphi(f_1),\ldots,\varphi(f_n)$
fulfilling all the Wightman axioms except by following modifications:
({\em a}) fields are operator-valued generalized functions
living in an appropriate space of functions $f(x) \in {\cal S}^0({\Bbb R}^{4n})$,
the space of entire analytic test functions, ({\em b}) the local commutativity
is replaced by the asymptotic variant in the sense of Soloviev, which
can be understood saying that two operators $\varphi(f)$ and $\varphi(g)$, at two
distinct points $x_1$ and $x_2$, can only be distinguished if the relative spatial
distance between the two points $x_1$ and $x_2$ is sufficiently large.

%%%%%%%%%%%%%%%%%%%%%%%%%%%%%%%%%%%%%%%%%%%%%%%%%%%%%%%%%%%%%%%%%%%%%
%                        POSTULATES
%%%%%%%%%%%%%%%%%%%%%%%%%%%%%%%%%%%%%%%%%%%%%%%%%%%%%%%%%%%%%%%%%%%%%

\,\,\,{\bf Fields are Analytical Functionals.}
The fields are operator-valued generalized functions on ${\cal S}^0({\Bbb R}^{4n})$,
the space of entire analytic test functions. In particular, we will consider only
one neutral scalar field $\varphi(x)$. We denote by $D_0$ the minimal common invariant
domain, which is assumed to be dense, of the field operators in the Hilbert space ${\mathscr H}$
of states, {\em i.e.}, the vector subspace of ${\mathscr H}$ that is spanned by the vacuum
state $| \Omega_o \rangle$ and by various vectors of the form
\[
\varphi(f_1)\star \dots \star \varphi(f_n)| \Omega_o \rangle=
e^{\frac{i}{2}\theta^{\mu\nu}
\sum\limits_{i<j}\frac{\partial}{\partial x_i^\mu}\frac{\partial}{\partial x_j^\nu}}
\varphi(f_1) \cdots \varphi(f_n)| \Omega_o \rangle,
\]
where $f_i(x_i) \in {\cal S}^0(\oR^4)$, {\em i.e.}, we axiomatize NCFT in the sense
of a field theory on a non-commutative space-time encoded by a Moyal product on the
test function algebra. It should be noted that, the space ${\cal S}^0$, being
Fourier-isomorphic to ${\mathscr D}$, is {\it nuclear}. Therefore, the $n$-point vacuum
expectation values uniquely determine Wightman generalized functions
${\mathscr W}_{\star}\in{\cal S}^{\prime 0}(\oR^{4n})$:
\begin{eqnarray}
{\mathscr W}_{\star}(f_{1} \otimes \cdots \otimes f_{n})\overset{\text{def}}
{=}\langle\Omega_o|\varphi(f_1)\star \cdots
\star \varphi(f_n)|\Omega_o\rangle\,.
\label{NWF}
\end{eqnarray}
Here $(f_{1} \otimes \cdots \otimes f_{n})(x_1,\ldots,x_n)=f_1(x_1)\cdots f_n(x_n)$
is considered as an element of ${\cal S}^0(\oR^{4n})$, where $\oR^{4n}$ is the
space of $n$-tuple of Minkowski vectors $(x_1,\ldots,x_n)$.
Eq.(\ref{NWF}) represents the Weyl form of the operator-valued Wightman functions
${\mathscr W}_n(f_{1} \otimes \cdots \otimes f_{n})$ proposed recently by
Chaichian {\it et al.}~\cite{Cha1} (apparently, a similar definition of
Wightman functions for the non-commutative case was used in~\cite{Maha}).
The relation between ${\mathscr W}_\star(f_1 \otimes \cdots \otimes f_n)$ defined above
and the ordinary Wightman functions ${\mathscr W}_n(f_1 \otimes \cdots \otimes f_n)$ is
given by
\begin{align*}
{\mathscr W}_\star(f_1 \otimes &\cdots \otimes f_n)=
e^{\frac{i}{2}\theta^{\mu\nu}\sum\limits_{i<j}\frac{\partial}{\partial x_i^\mu}
\frac{\partial}{\partial x_j^\nu}}{\mathscr W}_n(f_1 \otimes \cdots \otimes f_n).
\end{align*}
This is a consequence of the product
\begin{align}
\varphi(f_1)\star &\cdots \star \varphi(f_n)=
e^{\frac{i}{2}\theta^{\mu\nu}\sum\limits_{i<j}\frac{\partial}{\partial x_i^\mu}
\frac{\partial}{\partial x_j^\nu}}\varphi(f_1) \cdots \varphi(f_n)\,\,.
\label{Moyalprod}
\end{align}
For coinciding points $x_1=x_2=\cdots=x_n$ the product (\ref{Moyalprod}) becomes identical
to the multiple Moyal $\star$-product. The use of the new Wightman functions (\ref{NWF})
suggests that the corresponding QFT is genuinely a NCFT~\cite{Cha1}.

\,\,\,{\bf Asymptotic Commutativity.}
If $f$ and $g$ are two test functions in ${\cal S}^0({\Bbb R}^4)$, then
the fields $\varphi(f)$ and $\varphi(g)$ are said to commute
asymptotically for sufficiently large space-like separation of their
arguments, if the functional
\begin{align}
f&=\bigl\langle\Theta,\,\bigl[\varphi(f),\varphi(g)\bigr]_\star
\Psi\bigr\rangle \nonumber\\[3mm]
&=\bigl\langle\Theta,\,\bigl(\varphi(f) \star \varphi(g)-
\varphi(g) \star \varphi(f)\bigr)\Psi\bigr\rangle\,\,,
\label{AofAC}
\end{align}
is carried by the closed cone $\overline{V}_{e}^{\,2} \times \oR^4$,
for any vectors $\Theta,\Psi\in D_0$, where $\overline{V}_{e}^{\,2}=
\bigl\{(x_e,x_e^\prime) \in \oR^4 \mid (x_{e}-x_{e}^\prime)^2 \geq 0\bigr\}$. Here,
following the Ref.~\cite{AGVM}, we call the coordinates $x_e=(x^0,x^1)$ ``electrical''
coordinates and $\vec{x}_m=(x^2,x^3)$ ``magnetic'' coordinates. 

In~\cite{AGVM} the condition that the functional (\ref{AofAC}) is carried by
the closed cone $\overline{V}_e^{\,2} \times \oR^4$ has been supported by the
$SO(1,1)\times SO(2)$ symmetry, which is the feature arising when one has only spatial
non-commutativity. This fact leads the authors of~\cite{AGVM} to argue that
the notion of a light cone is generally modified to that of a light wedge. More recently,
Chu {\it et al.}~\cite{Chu} have shown that the reduction from light cone to light wedge
is a generic effect for non-commutative geometry and it is independent of the type of
Lorentz symmetry breaking interaction.

\begin{remark}
The analysis based in the symmetry $SO(1,1) \times SO(2)$ shows an
inconvenience: particles cannot be classified according to the 4-dimensional Wigner
particle concept. However, the recent work of Chaichian {\em et al}~\cite{Chai}
shows that this can be by passed invoking the concept of {\bf twisted} Poincar\'e
symmetry. From now on, we will assume that our fields are trans\-for\-ming according to
representations of the {\bf twisted} Poincar\'e group. This implies that the Wightman
functionals defined in (\ref{NWF}) satisfy the {\bf twisted} Poincar\'e covariance
condition.
\end{remark} 

In our arguments the equality ${\cal S}^0(\oR^{4}\times \oR^{4})={\cal S}^0(\oR^{4})
\widehat{\otimes}_i {\cal S}^0(\oR^{4})$ plays an important role, where the index $i$
indicates that the tensor product is endowed with the inductive topology and the hat means
the corresponding completion. By definition of the inductive topology, the dual space of
${\cal S}^0(\oR^{4}) \widehat{\otimes}_i {\cal S}^0(\oR^{4})$ is isomorphic to the
space of separately continuous functionals on ${\cal S}^0(\oR^{4}) \times 
{\cal S}^0(\oR^{4})$ (see~\cite{Solo3,Solo4}). By result $\bf R.2$ in Appendix,
the space ${\cal S}^0(\oR^{4})$ is dense in ${\cal S}^0(U)$, where $U$ is any open cone
in $\oR^{4}$ such that $\overline{V}_{e}^{\,2} \setminus \{0\} \subset U$. Hence,
the functional $f$ is carried by the closed cone $\overline{V}_{e}^{\,2} \times \oR^4$.
In addition, a consideration analogous to that of Lemma 3 in~\cite{Solo3} shows that if we
introduce, for $0 \leq j \leq n$ and $n=0,1,2,\ldots$, Wightman functions
${\mathscr W}_\star \in {\cal S}^{\prime 0}(\oR^{4(n+2)})$ defined by
\begin{align*}
{\mathscr W}_\star(f_1 \otimes \cdots \otimes f_j \otimes f &\otimes g
\otimes f_{j+1} \otimes \cdots \otimes f_n)\pm\\[3mm]
&{\mathscr W}_\star(f_1 \otimes \cdots \otimes f_j \otimes g \otimes f
\otimes f_{j+1} \otimes \cdots \otimes f_n)\,\,,
\end{align*}
it follows from the asymptotic commutativity condition that ${\mathscr W}_\star$
defined on ${\cal S}^{0}(\oR^{4(n+2)})$ has a continuous extension to the space
${\cal S}^{0}(\oR^{4j}\times (U \times \oR^4) \times \oR^{4(n-j)})$, where $U$ is
any open cone in $\oR^4$ such that $\overline{V}_{e}^{\,2} \setminus \{0\} \subset U$.
Then, ${\mathscr W}_\star$ is carried by the closed cone  
$\oR^{4j}\times (\overline{V}_{e}^{\,2} \times \oR^4) \times \oR^{4(n-j)}$.

\begin{remark}
From the estimates of the behavior of test functions belonging to space ${\cal S}^0$
over cones (see Eq.(\ref{eq.1})), the feature of the functional (\ref{AofAC}) to be
carried by $\overline{V}_{e}^{\,2}\times \oR^4$ is equivalent to the property of this
functional to decrease faster than any linear exponential function in the complement
of $\overline{V}_{e}^{\,2}\times \oR^4$. This statement is a special case
of Theorem 2 (Convolution Theorem) in~\cite{FaiSolo}. Particularly, this brings us back
to an old work due to Borchers-Pohlmeyer~\cite{BorPoh}, who considered the theory of
a {\it tempered} scalar field and replaced the microcausality condition by
an apparently weaker condition: they established that the bound of the form
\begin{align}
|\langle \Omega_o|[\varphi(x_1),\varphi(x_2)]\varphi(x_3) \cdots \varphi(x_n)|
\Omega_o \rangle| \le C_n e^{-\epsilon|(x_1-x_2)^2|^{\alpha/2}}\,,
\quad{\mbox{for}}\quad \alpha > 1\,,
\label{TBP}
\end{align}
on the behavior of the commutators at those points
$(x_1,x_2,\ldots,x_n)$ which belong to the open cone ${\mathscr J}_n$, the Jost
cone~\cite{CJost}, together with $(x_2,x_1,\ldots,x_n)$ results in the strict local
commutativity. In other words, they proved that a vanishing of the
commutator of the fields in the relative spatial distance $(x_1-x_2)$ which is
faster than a certain exponential decay leads back to a microlocal theory.
However, as observed by Soloviev~\cite{Solo3}, the asymptotic commutativity condition
being applied to the tempered field does not amount to the naive bound (\ref{TBP})
and does not imply local commutativity. The asymptotic commutativity condition
indicates a fast decrease not of the commutator itself but of the result of
smoothing it by convolution with appropriate test functions~\cite{FaiSolo}.
\end{remark}

%%%%%%%%%%%%%%%%%%%%%%%%%%%%%%%%%%%%%%%%%%%%%%%%%%%%%%%%%%%%%%%%%%%%%%%%%%%%
\section{The Notion of Equivalence Classes of a NC Field}
%\hspace*{\parindent}
%%%%%%%%%%%%%%%%%%%%%%%%%%%%%%%%%%%%%%%%%%%%%%%%%%%%%%%%%%%%%%%%%%%%%%%%%%%%
As it is well-known, the Borchers class of a quantum field is a direct
consequence of the CPT theorem. For this reason, we shall recall the validity
of this theorem within the context of a NCFT~\cite{Cha1,DCJhep}. For simplicity,
we will assume a NCFT given in terms of a single neutral scalar field, in the
case of only spatial non-commutativity.

\begin{theorem}[Modified CPT Theorem]
A non-commutative scalar field theory symmetric under the CPT-operation $\Theta$ is
equivalent to the weak asymptotic commutative condition. 
\label{cpttheo} \end{theorem}

To prove the Theorem \ref{cpttheo}, we must develop our machinery a little further.

\begin{proposition}
The functional ${\mathscr F}(f)={\mathscr W}_\star(f_1 \otimes \cdots \otimes f_n)-
{\mathscr W}_\star(\widehat{f}_1 \otimes \cdots \otimes \widehat{f}_n)$, where
$\widehat{f}_i(x_i)={f}_i(-x_i)$, is carried by the complement of the Jost real
points ${\mathscr J}_n$, characterized by the set
\begin{align*}
\Bigl\{(x_1,\ldots,x_n) \in \oR^{4n}  \bigm|
\Bigl(\sum_{i=1}^{n-1}\lambda_i(x_{e_i}-x_{e_{i+1}})\Bigr)^2 < 0 \Bigr\}\,\,,
\end{align*}
\label{auxpro}
for an arbitrary set of $\lambda_i$ satisfying the conditions $\lambda_i \geq 0$
and $\sum_{i=1}^{n-1}\lambda_i > 0$.
\end{proposition}

\begin{proof}
The proof follows using arguments paralleling the analysis of the proof of the
Proposition 4.1 in~\cite{DCJhep}, taking into account that the new Wightman functions
no affect the analytic continuation of these functions to the complex plane with
respect the ``electrical'' coordinates $x_e=(x^0,x^1)$.
\end{proof}

\begin{remark}
According to \'Alvarez-Gaum\'e and V\'asquez-Mozo~\cite{AGVM},
for $n>2$ the Jost points are formed by $(x_{e_i}-x_{e_{i+1}})^2<0$ with
the condition that $x_i^1-x_{i+1}^1>0$.
\end{remark}

\begin{definition}
The non-commutative quantum field $\varphi(x)$ defined on the test function space
${\cal S}^0({\oR^4})$ is said to satisfy the weak asymptotic commutativity (WAC)
condition if the functional 
\[
{\mathscr W}_\star(f_1 \otimes \cdots \otimes f_n)-
{\mathscr W}_\star(f_n \otimes \cdots \otimes f_1)\,\,,
\]
is carried by the set $\complement{\mathscr J}_n$ complementary to the Jost points.
\end{definition}

\begin{proof}[Proof of Theorem \ref{cpttheo}]
The CPT invariance condition is derived by requiring that the CPT
operator $\Theta$ be antiunitary -- see~\cite{SW,BLOT,Haag}:
\begin{equation}
\langle\Theta\Xi|\Theta\Psi\rangle=\langle\Psi|\Xi\rangle\,\,,
\label{antiunit}
\end{equation}
this means that the CPT operator leaves invariant all transition probabilities
of the theory. In the case of a NCFT, the operator $\Theta$ can be constructed in
the ordinary way. Taking the vector states as $\langle\Xi|=\langle\Omega_o|$ and
$|\Psi\rangle=\varphi(f_n)\star \cdots \star \varphi(f_1)|\Omega_o\rangle$
we shall express both sides of (\ref{antiunit}) in terms
of NC Wightman functions. For the left-hand side of (\ref{antiunit}) we can use directly
the CPT transformation properties of the field operators, which for a neutral scalar
field is equal to $\Theta \varphi(f)\Theta^{-1}=\varphi(\widehat{f})$. Using the
CPT-invariance of the vacuum state, $\Theta|\Omega_o\rangle=|\Omega_o\rangle$,
the left-hand side of (\ref{antiunit}) becomes:
\begin{align}
\langle\Theta\Xi|\Theta\Psi\rangle &=
\langle\Theta\Omega_o|\Theta(\varphi(f_n)\star \cdots \star
\varphi(f_1)|\Omega_o\rangle \nonumber\\[3mm]
&={\mathscr W}_\star(\widehat{f}_n \otimes \cdots \otimes \widehat{f}_1)\,\,.
\label{thetaw2}
\end{align}

In order to express the right-hand side of (\ref{antiunit}), we take the
Hermitian conjugates of the vectors $|\Psi\rangle$ and $\langle\Xi|$, to obtain:
\begin{equation}
\langle\Psi|\Xi\rangle={\mathscr W}_\star(f_1 \otimes \cdots \otimes f_n)\,\,.
\label{thetaw1}
\end{equation}
Putting together (\ref{antiunit}) with (\ref{thetaw2}) and (\ref{thetaw1}), we
obtain the CPT invariance condition in terms of NC Wightman functions as
\begin{equation}
{\mathscr W}_\star(f_1 \otimes \cdots \otimes f_n)=
{\mathscr W}_\star(\widehat{f}_n \otimes \cdots \otimes \widehat{f}_1)\,\,.
\label{cpt-wight}
\end{equation}
Now, we subtract the functional ${\mathscr W}_\star(\widehat{f}_1 \otimes
\cdots \otimes \widehat{f}_n)$ from the left-hand and right-hand sides of
(\ref{cpt-wight}) and obtain
the expression:
\begin{align*}
\Bigl[{\mathscr W}_\star(f_1 \otimes \cdots \otimes f_n)-
{\mathscr W}_\star(\widehat{f}_1 \otimes \cdots \otimes \widehat{f}_n)\Bigr]=
\Bigl[{\mathscr W}_\star(\widehat{f}_n \otimes \cdots \otimes \widehat{f}_1)-
{\mathscr W}_\star(\widehat{f}_1 \otimes \cdots \otimes \widehat{f}_n)\Bigr]\,.
\end{align*}
By Proposition \ref{auxpro}, the difference functional on the left-hand side, denoted by
${\mathscr F}(f)$, is carried by set $\complement{\mathscr J}_n$. It follows that
the functional on the right-hand side is also carried by this set. This means that the
WAC is fulfilled, taking into consideration the symmetry ${\mathscr J}_n=-{\mathscr J}_n$
of the Jost points. The reverse is also easily proved. If the WAC is satisfied, then the
difference ${\mathscr W}_\star(f_1 \otimes \cdots \otimes f_n)-
{\mathscr W}_\star(\widehat{f}_n \otimes \cdots \otimes \widehat{f}_1)$
is carried by $\complement{\mathscr J}_n \not= \oR^{2n} \times \oR^{2n}$. On the other hand,
by virtue of the spectral condition, the Fourier transform of this difference
has support in the properly convex cone
\[
\Bigm\{(p_1,\ldots,p_n)\in {\Bbb R}^{4n}\,\,\bigm|\,\,\sum_{j=1}^{n}p_j=0,\,\,
\sum_{j=1}^{k}p_j \in {\overline V}_+,\,\,k=1,\dots,n-1 \Bigm\}\,\,.
\]
Therefore, the CPT invariance holds identically by Theorem 4 in~\cite{Solo2}, which
asserts that ${\mathscr W}_\star(f_1 \otimes \cdots \otimes f_n)-
{\mathscr W}_\star(\widehat{f}_n \otimes \cdots \otimes \widehat{f}_1)
\equiv 0$, since the property of this functional of having its Fourier
transform supported by the aforementioned properly convex cone requires that each carrier
cone of ${\mathscr W}_\star(f_1 \otimes \cdots \otimes f_n)-
{\mathscr W}_\star(\widehat{f}_n \otimes \cdots \otimes \widehat{f}_1)$ cannot be
different from the whole space $\oR^{2n} \times \oR^{2n}$.
\end{proof}

Finally, the Borchers class of quantum fields for a NCFT is a consequence of the
following theorem.

\begin{theorem}
Suppose $\varphi(f)$ is a field satisfying the assumptions of Theorem \ref{cpttheo}
and $\Theta$ is the corresponding $CPT$-symmetry operator. Suppose $\psi(f)$ is
another field transforming under the same representation of the {\bf twisted}
Poincar\'e group, with the same domain of definition. Suppose that the functional
$\langle\Omega_o|\varphi(f_1)\star\cdots\star
\varphi(f_m)\star\psi(f)\star\varphi(f_{m+1})\star\cdots\star\varphi(f_n)
|\Omega_o\rangle - \langle\Omega_o|\varphi(f_n)\star\cdots\star\varphi(f_{m+1})\star
\psi(f)\star\varphi(f_m)\star\cdots\star\varphi(f_1)
|\Omega_o\rangle$ is carried by $\complement{\mathscr J}_{n+1}$.
Then $\Theta$ implements the $CPT$ symmetry for $\psi(f)$ as well and the fields 
$\varphi,\psi$ satisfy the weak asymptotic commutativity condition.
\label{Theo5}
\end{theorem}

\begin{proof}
Applying the same arguments as above, we find that
\begin{align}
\langle\Omega_o|\varphi(f_1)\star\cdots \star &\varphi(f_m)\star\psi(f)\star
\varphi(f_{m+1})\star\cdots\star\varphi(f_n)|\Omega_o\rangle= \nonumber\\[3mm]
&\langle\Omega_o|\varphi(\widehat{f}_n)\star\cdots\star\varphi(\widehat{f}_{m+1})
\star \psi(\widehat{f})\star\varphi(\widehat{f}_m)\star \cdots \star
\varphi(\widehat{f}_1)|\Omega_o\rangle\,\,,
\label{eqno24}
\end{align}

If $\Theta$ is the CPT operator for the field $\varphi$, it follows that for
any $\Psi$
\begin{equation}
\langle\Theta\Xi|\Theta\psi(f)\Theta\Theta^{-1}|\Psi\rangle=
\langle\overline{\Xi|\psi(f)|\Psi}\rangle\,\,,
\label{eqno26}
\end{equation}
and if
\[
|\Xi\rangle=\varphi(f_m)\star\cdots\star\varphi(f_1)|\Omega_o\rangle
\quad{\mbox{and}}\quad
|\Psi\rangle=\varphi(f_{m+1})\star\cdots\star\varphi(f_n)|\Omega_o\rangle\,\,,
\]
then
\[
|\Theta\Xi\rangle=\varphi(\widehat{f}_m)\star\cdots
\star\varphi(\widehat{f}_1)|\Omega_o\rangle
\quad{\mbox{and}}\quad
|\Theta\Psi\rangle=\varphi(\widehat{f}_{m+1})\star\cdots
\star\varphi(\widehat{f}_n)|\Omega_o\rangle\,\,.
\]
Using (\ref{eqno24}), we obtain
\[
\langle\Psi|\psi(f)|\Xi\rangle=
\langle\Theta\Xi|\psi(\widehat{f})\Theta|\Psi\rangle\,\,,
\]
which when compared with (\ref{eqno26}) leads to
\[
\Theta\,\psi(f)\,\Theta^{-1}=\psi(\widehat{f})\,\,.
\]
Thus, the operator $\Theta$ transforms $\psi$ correctly, as was to be proved.
\end{proof}

\begin{corollary}[Transitivity of Weak Relative Asymptotic Commutativity]
The weak relative asymptotic commutativity property is transitive in the sense
that if each of the fields $\psi_1,\psi_2$ satisfies the assumptions of
Theorem \ref{Theo5}, then there is a CPT-symmetry operator common to the fields
$\{\varphi,\psi_1,\psi_2\}$ and by Theorem \ref{cpttheo}, the weak relative
asymptotic commutativity condition is satisfied not only for $\{\psi_1,\psi_2\}$
but also for $\{\varphi,\psi_1,\psi_2\}$.
\end{corollary}

%%%%%%%%%%%%%%%%%%%%%%%%%%%%%%%%%%%%%%%%%%%%%%%%%%%%%%%%%%%%%%%%%%%
\section{Final Considerations}
%\hspace*{\parindent}
%%%%%%%%%%%%%%%%%%%%%%%%%%%%%%%%%%%%%%%%%%%%%%%%%%%%%%%%%%%%%%%%%%%
In this work, we continue applying some ideas by Soloviev~\cite{Solo1}-\cite{Solo4}
as an alternative
description towards an axiomatic formulation of non-commutative quantum field theory.
Following our previous paper~\cite{DCJhep}, we have extended the Borchers class of
quantum fields for a NCFT, by replacing the weak relative locality condition by
a weak relative asymptotic commutativity property. We restrict ourselves to the simplest
case of a neutral scalar field, in the case of spatial non-commutativity. An important
consequence of this result is that the Theorem 4.20 in~\cite{SW} on S-equivalence
of quantum fields can be generalized for NCFT, by adapting the Theorem 6 in~\cite{Solo4}
to the basic postulates of Section 2.
 
%%%%%%%%%%%%%%%%%%%%%%%%%%%%%%%%%%%%%%%%%%%%%%%%%%%%%%%%%%%%%%%%%%%%%%%%%%%
\section*{Acknowledgments}
%%%%%%%%%%%%%%%%%%%%%%%%%%%%%%%%%%%%%%%%%%%%%%%%%%%%%%%%%%%%%%%%%%%%%%%%%%%
I would like to thank Professor O. Piguet for his kind invitations
at the Departamento de F\'\i sica, Universidade Federal do Esp\'\i rito Santo
(UFES) and  C.P. Constantinidis for encouragement.
%%%%%%%%%%%%%%%%%%%%%%%%%%%%%%%%%%%%%%%%%%%%%%%%%%%%%%%%%%%%%%%%%%%%%%%%%%%%%
%%%%%%%%%%%%%%%%%%%%%%%%%%%%%%%%%%%%%%%%%%%%%%%%%%%%%%%%%%%%%%%%%%%%%%%%%%%%%
\appendix 
\renewcommand{\theequation}{\Alph{section}.\arabic{equation}}
\renewcommand{\thesection}{\Alph{section}}

\setcounter{equation}{0} \setcounter{section}{0}

%%%%%%%%%%%%%%%%%%%%%%%%%%%%%%%%%%%%%%%%%%%%%%%%%%%%%%%%%%%%%%%%%%%%%%%%%%%
\section{Angular Localizability}
\label{Ap1}
%\hspace*{\parindent}
%%%%%%%%%%%%%%%%%%%%%%%%%%%%%%%%%%%%%%%%%%%%%%%%%%%%%%%%%%%%%%%%%%%%%%%%%%%%
In this appendix, for sake of completeness, we recall some results by
Soloviev which make possible to extend the basic results of axiomatic
approach~\cite{SW,BLOT,Haag} to nonlocal interactions -- the reader is
referred to~\cite{Solo1}-\cite{Solo4} and references therein for details.

We start recalling that the space of test functions composed by entire analytic
functions is such that the following estimate holds:
\[
|f(z)| \le C_N\left(1+|x|\right)^{-N}e^{b|y|}\,\,,
\]
where $z=x+iy$ and $N \in {\Bbb N}$. $C_N$ and $b$ are positive constants depending
on $f$. The space of functions
satis\-fying the estimate above, with fixed $b$, is denoted as ${\cal S}^{0,b}$,
while in nonlocal field theory the union $\cup_{b>0}{\cal S}^{0,b}$ is denoted as
${\cal S}^0$. Together with space ${\cal S}^0(\oR^n)$, we introduce a space associated
with closed cones $K \subset \oR^n$. One recalls that $K$ is called a cone if $x \in K$
implies $\lambda x \in K$ for all $\lambda > 0$. Let $U$ be an open cone in $\oR^n$.
For each $U$, one assigns a space ${\cal S}^0(U)$ consisting of those entire analytic
functions on $\oC^n$, that satisfy the inequalities
\begin{align}
|f(z)| \le C_N\left(1+|x|\right)^{-N}e^{b|y|+ bd(x,U)}\,\,,
\label{eq.1}
\end{align}
with $d(x,U)$ being the distance from the point $x$ to the cone $U$. Here, the
norm in $\oR^n$ is assumed to be Euclidean. This space can naturally be given a
topology by regarding it as the inductive limit of the family of countably normed
spaces ${\cal S}^{0,b}(U)$ whose norms are defined in accordance with the inequalities
(\ref{eq.1}), {\em i.e.},
{\small
\[
\|f\|_{U,b,N} =
\sup_z |f(z)|\left(1+|x|\right)^N e^{-b|y|-bd(x,U)}.
\]}
For each closed cone $K\subset \oR^n$, one also defines a space ${\cal S}^0(K)$
by taking another inductive limit through those open cones $U$ that contain
the set $K\setminus\{0\}$ and shrink to it. Clearly, ${\cal S}^0(\oR^n)={\cal S}^0$.
As usual, we use a prime to denote the continuous dual of a space under consideration.
A closed cone $K\subset\oR^n$ is said to be a {\it carrier} of a  functional
$v\in {\cal S}^{\prime 0}$ if $v$ has a continuous extension to the space
${\cal S}^0(K)$, {\em i.e.}, belongs to ${\cal S}^{\prime 0}(K)$. As is seen from
estimate (\ref{eq.1}), this property may be thought of as a fast decrease -- no worse
than an exponential decrease of order 1 and maximum type -- of $v$ in the complement
of $K$. It should also be emphasized that  if  $v$ is a tempered distribution with
support in $K$, then the restriction $v|_{{\cal S}^0}$ is carried by $K$.

We now list some results proved by Soloviev, which formalize the property of angular
localizability:
%%%%%%%%%%%%%%%%%%%%%%%%%%%%%%%%%%%%%%%%%%%%%%%%%%%%%%%%%%%%%%%%%%%%%%%%%%%%%%%%%%%%%%%%%
\newcounter{numero}
\setcounter{numero}{0}
\def\Theo{\addtocounter{numero}{1}\item[{$\bf R.\thenumero$}]}

\begin{enumerate}

\Theo The spaces ${\cal S}^0(U)$ are Hausdorff and complete.
A  set $B\subset {\cal S}^0(U)$ is bounded if, and only if, it is contained
in some space ${\cal S}^{0,b}(U)$ and is bounded in each of its norms.

\Theo The space ${\cal S}^0$ is dense in every ${\cal S}^0(U)$ and in every
${\cal S}^0(K)$.

\Theo If a functional $v\in {\cal S}^{\prime 0}$ is carried by each of closed
cones $K_1$ and $K_2$, then it is carried by their intersection.

\Theo If $v\in {\cal S}^{\prime 0}(K_1 \cup K_2)$, then $v=v_1+v_2$, where
$v_j\in {\cal S}^{\prime 0}(U_j)$ and $U_j$ are any open cones such that
$U_j\supset K_j\setminus\{0\}$, $j=1,2$.
\end{enumerate}

%%%%%%%%%%%%%%%%%%%%%%%%%%%%%%%%%%%%%%%%%%%%%%%%%%%%%%%%%%%%%%%%%%%%%%%%%%%

%%%%%%%%%%%%%%%%%%%%%%%%%%%%%%%%%%%%%%%%%%%%%%%%%%%%%%%%%%%%%%%%%%%%%%%%%%%%%%
%%%%%%%%%%%%%%%%%%%%%%%%%%%%%%%%%%%%%%%%%%%%%%%%%%%%%%%%%%%%%%%%%%%%%%%%%%%%%%

\begin{thebibliography}{99}

\bibitem{AGVM} L. \'Alvarez-Gaum\'e and M.A. V\'azquez-Mozo, ``{\em General properties
of non-commutative field theories,}'' {\bf Nucl.Phys.} {\bf B668} (2003) 293.

\bibitem{Cha1} M. Chaichian, M.N. Mnatsakanova, K. Nishijima, A. Tureanu and
Yu. S. Vernov ``{\em Towards an axiomatic formulation of noncommutative quantum
field theory,}'' hep-th/0402212.

\bibitem{DCJhep} D.H.T. Franco and C.M.M. Polito, ``{\em A New Derivation of the
CPT and Spin-Statistics Theorems in Non-Commutative Field Theories,}'' hep-th/0403028,
revised version to appear in J.Math.Phys.

\bibitem{Luc1} W. L\"ucke, ``{\em PCT, spin and statistics, and all that for nonlocal
Wightman fields,}'' {\bf Commun.Math.Phys.} {\bf 65} (1979) 77.

\bibitem{Luc2} W. L\"ucke, ``{\em Spin-statistics theorem for fields with
arbitrary high energy behavior,}'' {\bf Acta Phys.Austr.} {\bf 55} (1984) 213.

\bibitem{Luc3} W. L\"ucke, ``{\em PCT theorem for fields with
arbitrary high-energy behavior,}'' {\bf J.Math.Phys.} {\bf 27} (1985) 1901.

\bibitem{Solo1} M.A. Soloviev, ``{\em Extension of the spin-statistics
theorem to nonlocal fields,}'' {\bf JETP} {\bf 67} (1998) 621.

\bibitem{Solo2} M.A. Soloviev, ``{\em A uniqueness theorem for distributions and its
application to nonlocal quantum field theory,}'' {\bf J.Math.Phys.} {\bf 39} (1998)
2635.

\bibitem{Solo3} M.A. Soloviev, ``{\em PCT, spin and statistics and analytic wave front
set,}'' {\bf  Theor.Math.Phys.} {\bf 121} (1999) 1377.

\bibitem{Solo4} M.A. Soloviev, ``{\em Nonlocal extension of the Borchers
classes of quantum fields,}'' in Multiple Facets of Quantization and Supersymmetry,
Ed. M. Olshanetsky and A. Vainshtein, contribution to Marinov Memorial Volume,
World Scientific, 697-717.

\bibitem{GeSh} I.M. Gelfand and G.E. Shilov, ``{\em Generalized functions,}'' Vol.2,
Academic Press Inc., New York, 1968.

\bibitem{Bor} H.-J. Borchers, ``{\em \"Uber die mannigfaltigkeit der
interpolierenden felder zu einer kausalen S-matrix,}''
{\bf Nuovo Cimento} {\bf 15} (1960) 784.

\bibitem{BEC} The equivalence class of a free Hermitian scalar field
was determined independently by B. Schroer (unpublished preprint) and
Epstein~\cite{HEps}. Epstein and Schroer's theorem states that the
Borchers class of the free field is made up with the set of Wick polynomials
of the free field and its derivatives defined as limits of Wick polynomials,
when all points are made to coincide:
\begin{align*}
\lim_{x_1,\ldots,x_n \rightarrow x}\wick{D^{\alpha_1}\varphi(x_1)\cdots
D^{\alpha_n}\varphi(x_n)}=
\wick{D^{\alpha_1}\varphi(x)\cdots
D^{\alpha_n}\varphi(x)}\,\,.
\end{align*}
In the case of a nonlocalizable field theory, the Borchers class was first
raised in~\cite{Consta}.

\bibitem{HEps} H. Epstein, ``{\em On the Borchers class of a free field,}''
{\bf Nuovo Cimento} {\bf 27} (1963) 886.

\bibitem{Consta} J.G. Taylor and F. Cosntantinescu, ``{\em Equivalence between
non-localizable and local fields,}'' {\bf Commun.Math.Phys.} {\bf 30} (1973) 211.

\bibitem{Maha} N. Mahajan, ``{\em PCT theorem in field theory on non-commutative space,}''
{\bf Phys.Lett.} {\bf B569} (2003) 85.

\bibitem{Chu} Chong-Sun Chu, K. Furuta and T. Inami ``{\em Locality, Causality and
Noncommutative Geometry,}'' hep-th/0502012.

\bibitem{Chai} M. Chaichian, P.P. Kulish, K. Nishijima and A. Tureanu,
``{\em On a Lorentz-Ivariant Interpretation of Noncommutative Space-Time and
Its Implications on Noncommutative QFT,}'' {\bf Phys.Lett.} {\bf B604} (2004) 98.

\bibitem{FaiSolo} V.Ya. Fainberg and M.A. Soloviev ``{\em Non-localizability and
asymptotic commutativity,}'' {\bf Theor.Math.Phys.} {\bf 93} (1992) 1438.

\bibitem{BorPoh} H.-J. Borchers and K. Pohlmeyer, ``{\em Eine scheinbare
abschw\"achung der lokalit\"atsbedingung. II,}'' {\bf Commun.Math.Phys.} {\bf 8} (1968)
269.

\bibitem{CJost} We recall that the Wightman functionals can be regarded as
boundary values of certain holomorphic functions~\cite{SW,BLOT,Haag}, with
the domain of analyticity having the form of an open tube, i.e.,
$T_{n-1}=\bigl\{(z_1,\ldots,z_n) \in \oC^{4n} \mid {\rm Re}\,z_i \in \oR^{4n},
{\rm Im}\,z_i \in V_+\bigr\}$. Here $V_+$ is the forward light cone in
$\oR^{4n}$, and for this reason $T_{n-1}$ is called the forward tube. $T_{n-1}$
has no real points, since the Wightman functionals are analytic only
when ${\rm Im}\,z_i \not= 0$. An elegant technical result,
the Bargamann-Hall-Wightman theorem, states the Wightman functionals can
be analytically continued into the extended tube $T_{n-1}^{\rm ext}$,
called the {\bf extended forward tube}, with the functionals being
covariant under the complex Lorentz group ${\mathscr L}_+(\oC)$.
$T_{n-1}^{\rm ext}$ contains certain real points, the so called
{\bf Jost points}, $\mathscr J_n$, which lie {\bf outside} the light cone
-- R. Jost, ``{\em Eine Bemerkung zum CPT,}'' {\bf Helv.Phys. Acta} {\bf 30}
(1957) 409.

\bibitem{SW} R.F. Streater and A.S. Wightman, ``{\em PCT, spin and statistics, and
all that,}'' Addison--Wesley, Redwood City, 1989.

\bibitem{BLOT} N.N. Bogoliubov, A.A. Logunov,  A.I. Oksak and I.T. Todorov,
``{\em General principles of quantum field theory,}'' Kluwer, Dordrecht, 1990.

\bibitem{Haag} R. Haag, ``{\em Local Quantum Physics,}'' Second Edition, Springer,
Berlin, 1996.

\end{thebibliography}
\end{document}